\documentclass{PoS}
            
\usepackage{psfig,colordvi,amssymb,amsmath,float}
\usepackage{graphicx} 
\usepackage{latexsym}                            
\usepackage{amsfonts}                            
\usepackage[mathscr]{eucal}                      
\usepackage{dcolumn}                             

\usepackage{slashed}
\usepackage{dsfont}                              

\usepackage{multirow}

\usepackage{bm}                                  

\newcommand{\be}{\begin{equation}}
\newcommand{\ee}{\end{equation}}
\newcommand{\bea}{\begin{eqnarray}}
\newcommand{\eea}{\end{eqnarray}}
\usepackage{slashed}
\usepackage{dsfont}
\usepackage[latin1]{inputenc}

\newcommand{\Op}{\mathcal{O}} 

\newcommand{\J}{\mathcal{J}} 

\def\eq#1{\rm Eq.~(\ref{#1})}
\def\lsim{\mathrel{\rlap{\lower4pt\hbox{\hskip1pt$\sim$}}
    \raise1pt\hbox{$<$}}}                
\def\gsim{\mathrel{\rlap{\lower4pt\hbox{\hskip1pt$\sim$}}
    \raise1pt\hbox{$>$}}}                

 \newcommand{\csw}{\, c_{\rm SW}}  
   
\newcommand{\Dlr}{\buildrel \leftrightarrow \over D\raise-1pt\hbox{}}
 \newcommand{\Dl}{\buildrel \leftarrow \over D\raise-1pt\hbox{}}
\newcommand{\Dr}{\buildrel \rightarrow \over D\raise-1pt\hbox{}}

\title{$\quad$ Renormalization constants for one-derivative\\ 
$\qquad$ fermion operators in twisted mass QCD
  \begin{center}
    \includegraphics[scale=0.12]{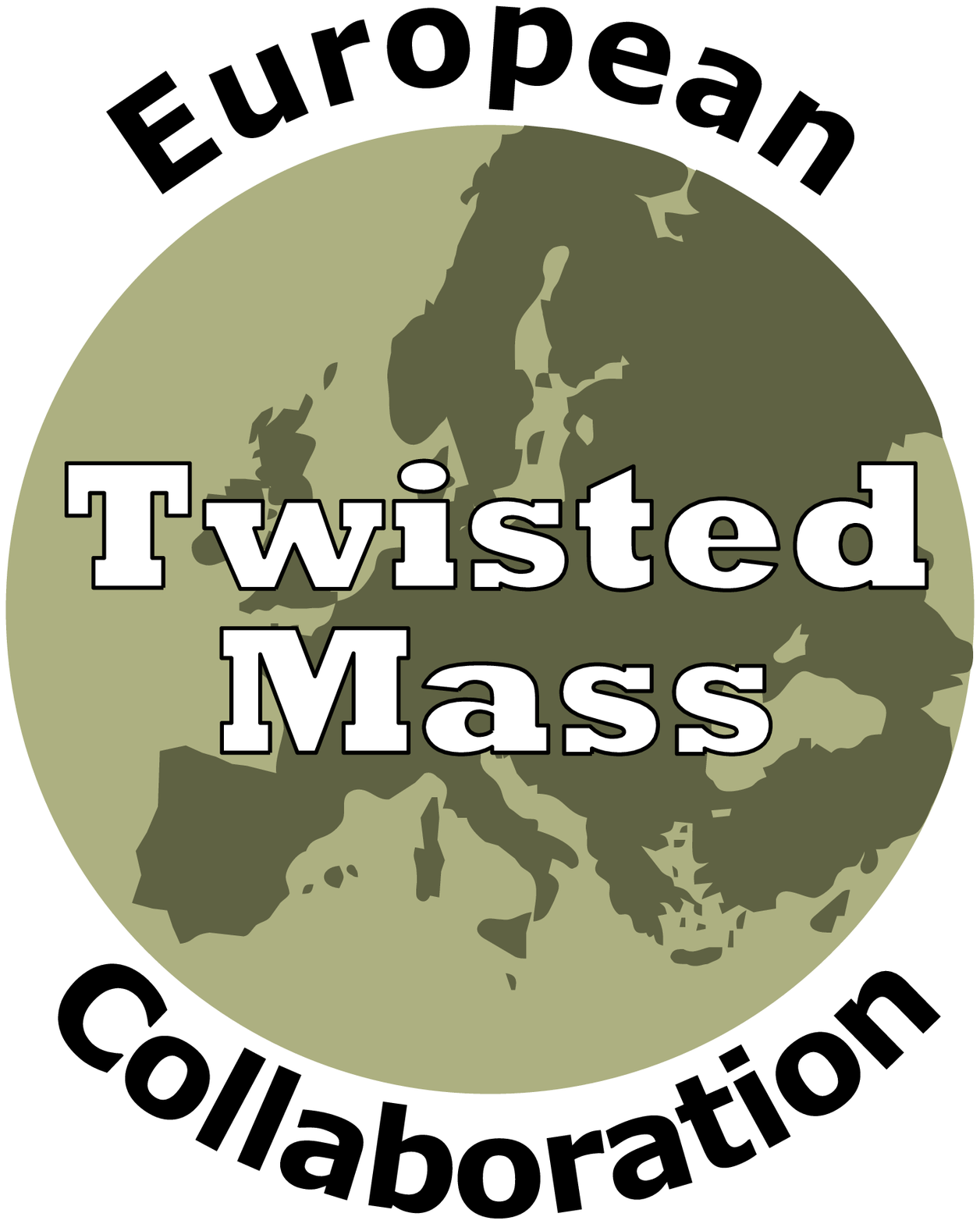}
  \end{center}}

\ShortTitle{Renormalization constants for one-derivative fermion operators in twisted mass QCD}

\author{C. Alexandrou\\
        Physics Department, University of Cyprus, The Cyprus Institute\\
        E-mail: \email{alexand@ucy.ac.cy}}
\author{\speaker{M. Constantinou}%
        \thanks{Work partly funded by the Cyprus Research Promotion Foundation: TECHNOLOGY/$\Theta$E$\Pi$I$\Sigma$/0308(BE)/17.} \\
       Physics Department, University of Cyprus\\
       E-mail: \email{marthac@ucy.ac.cy}}
\author{T. Korzec\\
        Institut f\"ur Physik,
   Humboldt Universit\"at zu Berlin\\
        E-mail: \email{korzec@physik.hu-berlin.de}}
\author{H. Panagopoulos\\
        Physics Department, University of Cyprus\\
        E-mail: \email{haris@ucy.ac.cy}}
\author{F. Stylianou\\
        Physics Department, University of Cyprus\\
        E-mail: \email{fstyl01@ucy.ac.cy}}

\abstract{We present perturbative and non-perturbative results on the
  renormalization constants of the local and one-derivative vector and axial vector
  operators. Non-perturbative results are obtained using the twisted
  mass Wilson fermion formulation employing two degenerate dynamical
  quarks and the tree-level Symanzik improved gluon action for pion
  masses in the range of about 450-260 MeV and at there values of the
  lattice spacing, namely 0.055 fm, 0.070 fm and 0.089 fm. Subtraction of ${\cal
    O}(a^2)$ terms is carried out by performing the perturbative
  evaluation of these operators at 1-loop and up to ${\cal
    O}(a^2)$. The renormalization conditions are defined in the
  RI$'$-MOM scheme, for both perturbative and non-perturbative
  results. The Z-factors, obtained for different values of the
  renormalization scale, are evolved perturbatively to a reference
  scale set by the inverse of the lattice spacing. In addition, they
  are translated to $\overline{MS}$ at 2 GeV using 3-loop perturbative results
  for the conversion factors.} 

\FullConference{The XXVIII International Symposium on Lattice Filed Theory\\
                June 14-19,2010\\
                Villasimius, Sardinia Italy}
\begin{document}

\section{Introduction}
Simulations in lattice QCD have advanced remarkably in the past
couple of years reaching the physical pion mass.
The theoretical and algorithmic improvements, combined with the tremendous increase in
computational power, have made {\it ab initio} calculations of
key observables on hadron structure in the chiral regime feasible enabling
comparison with experiment. Form factors and generalized parton
distribution functions (GPDs) can be obtained from the generalized form factors in
certain limiting cases. GPDs provide detailed information on the
internal structure of hadrons in terms of both the longitudinal
momentum fraction and the total momentum transfer squared. Beyond the
information that the form factors yield, such as size, magnetization
and shape, GPDs encode additional information, relevant for experimental
investigations, such as the decomposition of the total hadron spin
into angular momentum and spin carried by quarks and gluons.
GPDs are single particle matrix elements of the light-cone
operator~\cite{Ji:1998pc, Hagler:2003jd},
which can be expanded in terms of local twist-two operators
${\cal O}_\Gamma^{f,\{\mu_1\mu_2\cdots\mu_n\}}= \overline{\psi}^f\Gamma^{\{\mu_1}i\Dlr^{\mu_2}\cdots i\Dlr^{\mu_n\}}\psi^f$.
Lattice QCD allows us to extract hadron matrix elements for the
twist-2 operators, which can be expressed in terms of generalized form
factors.

In order to compare hadron matrix elements of these local operators to
experiment one needs to renormalize them. The aim of this paper is to
calculate non-perturbatively the renormalization factors of the above
twist-two fermion operators within the twisted mass formulation. We
show that, although the lattice spacings considered in this work are
smaller than $1$~fm, ${\cal O}(a^2)$ terms are non-negligible and
introduce significantly larger uncairtainties than statistical errors. We therefore compute the
${\cal O}(a^2)$ terms perturbatively and subtract them from the
non-perturbative results. This subtraction suppresses lattice
artifacts considerably depending on the operator under study and leads
to a more accurate determination of the renormalization constants~\cite{ACKPF1, ACKPF2}.

\section{Formulation}

For the gauge fields we use the tree-level Symanzik improved
gauge action~\cite{Weisz:1982zw}, which includes besides the
plaquette term also rectangular $(1\times2)$ Wilson loops.
The fermionic action for two degenerate flavors of quarks
 in twisted mass QCD is given by
\vspace{-0.35cm}
\be
S_F= a^4\sum_x  \overline{\chi}(x)\bigl(D_W[U] + m_0 
+ i \mu_0 \gamma_5\tau^3  \bigr ) \chi(x)
\label{action}
\ee
\vskip -0.35cm
\noindent with $\tau^3$ the Pauli matrix, $\mu_0$ the bare twisted mass 
and $D_W$ the massless Wilson-Dirac operator. Maximally twisted Wilson
quarks are obtained by setting the untwisted bare quark mass $m_0$ to
its critical value $m_{\rm cr}$, while the twisted quark mass
parameter $\mu_0$ is kept non-vanishing in order to give the light
quarks their mass. In $\eq{action}$ the quark fields $\chi$ are in the
so-called ``twisted basis''. The ``physical basis'' is obtained for
maximal twist by the simple transformations 
$\psi(x)=\exp\left(\frac {i\pi} 4\gamma_5\tau^3\right) \chi(x),\,\,
\overline\psi(x)=\overline\chi(x) \exp\left(\frac {i\pi} 4\gamma_5\tau^3\right)$.

Here we consider only the vector and axial twist-two operators up to
one-derivative, $Z_{\rm V}$, $Z_{\rm A}$, $Z_{\rm DV}$, $Z_{\rm DA}$
(symmetrized over two Lorentz indices and traceless), which are given
in the twisted basis as follows: 

{\scriptsize{
\begin{minipage}[h]{1.35in}
\vspace{-.85cm}
\begin{eqnarray}
   \Op_V^a \hspace{-0.1cm}&= \bar \chi \gamma_\mu\tau^a \chi &= \begin{cases} \bar \psi  \gamma_5\gamma_\mu \tau^2 \psi   & a=1 \\
                                                              -\bar \psi  \gamma_5\gamma_\mu \tau^1 \psi   & a=2 \\
                                                               \bar \psi  \gamma_\mu         \tau^3 \psi   & a=3 \nonumber \end{cases} \\
   \Op_A^a \hspace{-0.1cm}&= \bar \chi \gamma_5\gamma_\mu\tau^a \chi &= \begin{cases} \bar \psi  \gamma_\mu        \tau^2 \psi   & a=1 \\
                                                                      -\bar \psi  \gamma_\mu        \tau^1 \psi   & a=2 \\
                                                                       \bar \psi  \gamma_5\gamma_\mu\tau^3 \psi   & a=3 \nonumber \end{cases}
\end{eqnarray}
\end{minipage}
\hfill
\begin{minipage}[h]{3.5in}
\vspace{-.85cm}
\begin{eqnarray}
\Op_{\rm DV}^{\{\mu\,\nu\}\,a} \hspace{-0.2cm}&= \overline \chi \gamma_{\{\mu}\overleftrightarrow D_{\nu\}}\tau^a \chi 
                                              &= \begin{cases} \overline \psi  \gamma_5\gamma_{\{\mu}\overleftrightarrow D_{\nu\}} \tau^2 \psi   & a=1 \\
                                                              -\overline \psi  \gamma_5\gamma_{\{\mu}\overleftrightarrow D_{\nu\}} \tau^1 \psi   & a=2 \\
                                                               \overline \psi  \gamma_{\{\mu}\overleftrightarrow D_{\nu\}}         \tau^3 \psi   & a=3 \nonumber\end{cases} \\[0.5ex]
\Op_{\rm DA}^{\{\mu\,\nu\}\,a} \hspace{-0.2cm}&= \overline \chi \gamma_5\gamma_{\{\mu}\overleftrightarrow D_{\nu\}}\tau^a \chi 
                                              &= \begin{cases} \overline \psi  \gamma_{\{\mu}\overleftrightarrow D_{\nu\}} \tau^2 \psi   &\quad a=1 \\
                                                              -\overline \psi  \gamma_{\{\mu}\overleftrightarrow D_{\nu\}} \tau^1 \psi   & \quad a=2 \\
                                                               \overline \psi  \gamma_5\gamma_{\{\mu}\overleftrightarrow D_{\nu\}} \tau^3 \psi   & \quad a=3 \end{cases}
\end{eqnarray}
\end{minipage}
}}

\noindent In a massless
renormalization scheme the renormalization constants are defined in
the chiral limit, where isospin symmetry is exact. Hence, the same
value for $Z$ is obtained independently of the value of the isospin
index $a$ and therefore we drop the $a$ index from
here on. However, one must note that, for instance, the physical 
$\overline\psi \gamma_{\{\mu}\overleftrightarrow D_{\nu\}} \tau^1 \psi$ 
is renormalized with $Z_{\rm DA}$, while 
$\overline\psi \gamma_{\{\mu}\overleftrightarrow D_{\nu\}}\tau^3 \psi$
requires the $Z_{\rm DV}$, which differ from each other even in the chiral limit.
The one-derivative operators fall into different irreducible
representations of the hypercubic group, depending on the choice of
indices. Hence, we distinguish between
$\Op_{\rm DV1}\,(\Op_{\rm DA1}) = \Op_{\rm DV}\,\,(\Op_{\rm DA})$ with
$\mu=\nu$ and $\Op_{\rm DV2}\,(\Op_{\rm DA2}) = \Op_{\rm
  DV}\,(\Op_{\rm DA})$ with $\mu\neq\nu$.

\subsection{Renormalization Condition}

The renormalization constants are computed both perturbatively and
non-perturbatively in the RI$'$-MOM scheme at various renormalization
scales. We translate them to the ${\overline{\rm MS}}$-scheme at
(2~GeV)$^2$ using a conversion factor computed in perturbation theory
to ${\cal O}(g^6)$ as described in Section 3. The Z-factors
are determined by imposing the following conditions: 
{\small{
\be
   Z_q = \frac{1}{12} {\rm Tr} \left[\frac{-i \sum_\rho
     \gamma_\rho p_\rho}{p^2}\,(S^L(p))^{-1}\right] \Bigr|_{p^2=\mu^2} \,,\quad
   Z_q^{-1}\,Z^{\mu\nu}_{\cal O}\,\frac{1}{12} {\rm Tr} \left[(-i\,
     {\cal \tilde O}_{\{\mu} \,\, p_{\nu\}})^{-1}
     \,\Gamma^L_{\mu\nu}(p)\right] \Bigr|_{p^2=\mu^2} = 1\, ,
\label{renormalization cond}
\ee}}
\noindent where $\mu$ is the renormalization scale, $S_L$ and
$\Gamma_L$ correspond to the perturbative or non-perturbative
results and ${\cal \tilde O}_{\{\mu} \,\, p_{\nu\}}$ is the tree-level
expression of the operator under study. The trace is taken over spin
and color indices, and the conditions are imposed in the massless
theory.

\subsection{Perturbative procedure}

Our calculation for the Z-factors is performed in 1-loop perturbation
theory to ${\cal O}(a^2)$. The order $a^2$-terms can be subtracted
from non-perturbative estimates, and they can eliminate possible large
lattice artifacts. There are many difficulties when extracting powers
of the lattice spacing from our expressions, since there appear
singularities encountered at ${\cal O}(a^2)$, that persist even up to
6 dimensions (integral convergence in 7-d), making their extraction more delicate. In addition to
that, there appear Lorentz non-invariant contributions in ${\cal O}(a^2)$-terms,
such as $\sum_\mu p_\mu^4/p^2$, where $p$ is the external momentum; as
a consequence, the Z-factors also depend on such terms.

\begin{figure}[h]
\centerline{\psfig{figure=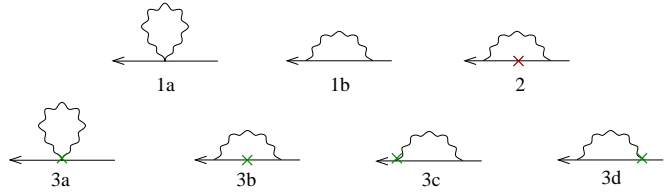,height=2.45truecm}}
\caption{One-loop diagrams contributing to the correction of the
amputated Green's functions of the propagator (1a, 1b), local bilinears
(2) and one-derivative operators (3a-3d). A wavy (solid) line represents gluons
(fermions). A cross denotes an insertion of the operator under study.}
\label{fig1}
\end{figure}

For all our perturbative results we employ a Wilson-type fermion action
(Wilson/clover/twisted mass), with non-zero bare mass, $m$.
For the renormalization of the fermion field and the local
bilinears we also have a finite twisted mass parameter, $\mu_0$, so we
can explore the mass dependence. For gluons we use Symanzik improved
actions (Plaquette, Tree-level Symanzik, Iwasaki, TILW, DBW2)
\cite{Constantinou:2009tr}.
The expressions for the matrix elements and the Z-factors are given in
a general covariant gauge, and their dependence on the coupling
constant, the external momentum, the masses and the clover parameter
$\csw$ is shown explicitly. The Feynman diagrams involved in the computation
of the various Z-factors are illustrated in Fig. \ref{fig1}. 
Here we do not show any expressions for the matrix elements of the
Green's functions, since they are far too lengthy. As an example we
show the ${\cal O}(a^2)$ terms that can improve the non-perturbative
estimate of $Z_q$ once they are subtracted. For the special choices:
$c_{SW}=0$, $r=1$ (Wilson parameter), $\lambda=0$ (Landau gauge),
$m_0=0$, $\mu_0=0$, and for tree-level Symanzik gluons, $Z_q$ can be
corrected to ${\cal O}(a^2)$ as follows:
%
\be
Z_q^{\rm impr}=Z_q^{\rm non-pert} - 
\frac{\Red{a^2}g^2C_F}{16\pi^2} \Big[\mu^2 \big( 1.1472 {-}\frac{73}{360}
  \ln(\Red{a^2}\mu^2) \big) {+}\frac{\sum_\rho \mu_\rho^4}{\mu^2} \big(2.1065 {-}\frac{157}{180} \ln(\Red{a^2}\mu^2) \big)
\Bigr]
\ee
Its most general expression is far too lengthy to be included in paper
form; it is provided, along with the rest of our results for the
Z-factors, in electronic form in Ref.~\cite{ACKPF2}.

\subsection{Non-perturbative calculation}

For each operator we define a bare vertex function given by
\vspace{-0.15cm}
\be\label{vfun}  
G(p) = \frac{a^{12}}{V}\sum_{x,y,z,z'} e^{-ip(x-y)} \langle u(x) \overline u(z) \J(z,z') d(z') \overline d(y) \rangle \, ,
\vspace{-0.15cm}
\ee
where $p$ is a momentum allowed by the boundary conditions, $V$ is the lattice volume, and the gauge average is
performed over gauge-fixed configurations. The form of $\J(z,z')$
depends on the operator under study, for example $\J(z,z') {=} \delta_{z,z'}
\gamma_\mu$ would correspond to the local vector current. In the
literature there are two main approaches that have been employed for
the evaluation of Eq.~(\ref{vfun}). The first approach relies on
translation invariance to shift the coordinates of the correlators in
Eq.~(\ref{vfun}) to position $z{=}0$ \cite{Constantinou:2010gr}. Having
shifted to $z{=}0$ allows one to calculate the
amputated vertex function for a given operator $\J$ for {\it any}
momentum with one inversion per quark flavor. In this work we explore
the second approach, introduced in Ref.~\cite{Gockeler:1998ye}, which
uses directly Eq.~(\ref{vfun}) without employing translation
invariance. One must now use a source that is momentum dependent but
can couple to any operator. For twisted mass fermions, with twelve
inversions one can extract the vertex function for a {\em single}
momentum. The advantage of this approach is a high statistical
accuracy and the evaluation of the vertex for any operator including
extended operators at no significant additional computational
cost. We fix to Landau gauge using a stochastic over-relaxation
algorithm~\cite{deForcrand:1989im}.

\section{Results}
\begin{figure}[h]
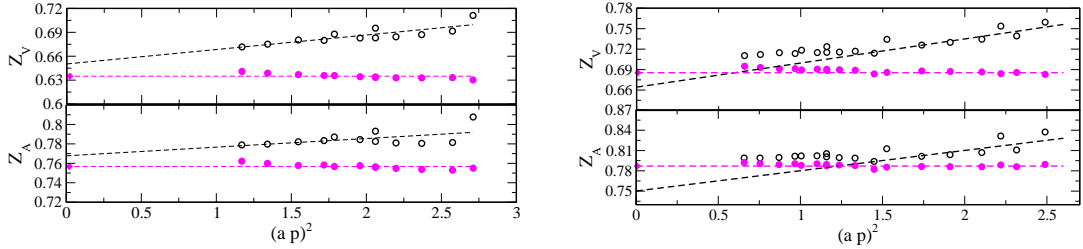

\centerline{\psfig{figure=Z_local_MSbar_2GeV_in_GeV_b3.9_PoS.eps,height=3.23truecm,clip=}$\qquad$\psfig{figure=Z_local_MSbar_2GeV_in_GeV_b4.20_PoS.eps,height=3.23truecm,clip=}}
\caption{Renormalization scale dependence for $Z_{\rm V}$, $Z_{\rm A}$ at
  $\beta=3.9,\,m_\pi=0.430$~GeV (left panel) and
  $\beta=4.20,\,m_\pi=0.476$~GeV (right panel) (Open points:
  unsubtracted, filled points: subtracted).}
\label{fig2}
\end{figure}

\begin{figure}[h]
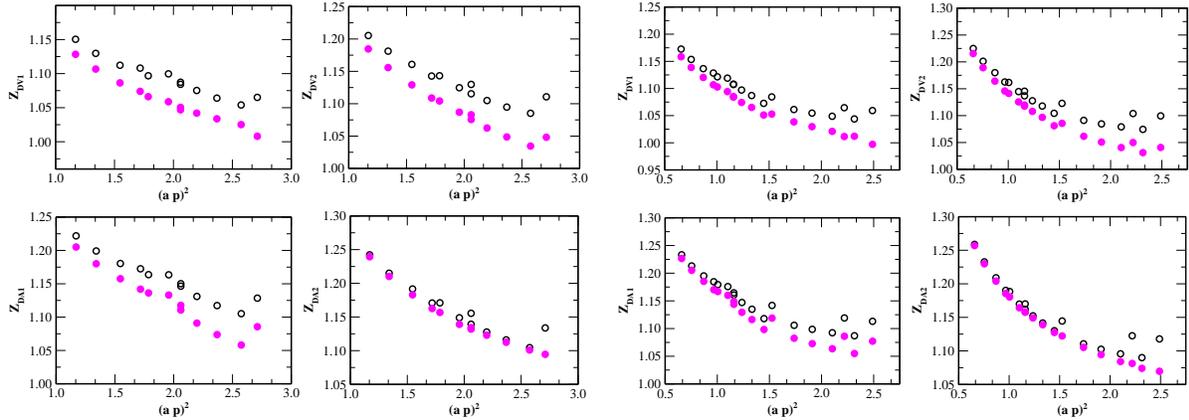

\centerline{\psfig{figure=Z_oneD_b3.9_0.0085_PoS.eps,height=5.55truecm,clip=}$\quad$\psfig{figure=Z_oneD_b4.20_0.0065_PoS.eps,height=5.55truecm,clip=}}
\caption{Renormalization scale dependence for $Z_{\rm DV}$, $Z_{\rm
    DA}$ (RI$'$-MOM scheme before evolving at a reference scale) at
  $\beta=3.9,\,m_\pi=0.430$~GeV (left panel) and
  $\beta=4.20,\,m_\pi=0.476$~GeV (right panel) (Open points:
  unsubtracted, filled points: subtracted).} 
\label{fig3}
\end{figure}

We perform the non-perturbative calculation of renormalization
constants for three values of the lattice spacing, $a$=0.089 fm, 0.070
fm, 0.056 fm, corresponding to $\beta=3.9,\,4.05$ and $4.20$
respectively. In Tables I and II of Ref.~\cite{ACKPF1} we summarize the
various parameters that we used in our simulations. We have tested
finite volume effects and pion mass dependence; both effects are
within the small statistical errors for the operators considered here.   
Chiral extrapolations are necessary to obtain the renormalization
factors in the chiral limit. Since the dependence on the pion mass is
insignificant, even if we allow a slope and perform a linear
extrapolation to our data, this is consistent with zero;
therefore the renormalization constants are computed at one quark
mass. Figures \ref{fig2}-\ref{fig3} demonstrate the effect of
subtraction at two $\beta$ values for the local and one-derivative
vector/axial Z-factors, as a function of the renormalization scale (in
lattice units). $Z_{\rm V}$ and $Z_{\rm A}$ are scale independent,
thus we obtain a very good plateau upon subtraction of ${\cal O}(a^2)$
effects. To identify a plateau for $Z_{\rm DV}$ and $Z_{\rm DA}$ we
need to convert to $\overline{\rm MS}$ and evolve to a reference scale.

\noindent{\bf{$\bullet$ Conversion to ${\overline{\rm MS}}$:}}
The passage to the continuum ${\overline{\rm MS}}$-scheme is
accomplished through use of a conversion factor, which is computed up
to 3 loops in perturbation theory. By definition, this conversion
factor is the same for the one-derivative vector and axial renormalization
constant, but will differ for the cases $Z_{\rm DV1}\,(Z_{\rm DA1})$
and $Z_{\rm DV2}\,(Z_{\rm DA2})$, that is 
$C_{\rm DV1} \equiv C_{\rm DA1} =Z^{\overline{\rm MS}}_{\rm DV}/Z^{\rm RI'}_{\rm DV1},\,
C_{\rm DV2} \equiv C_{\rm DA2} =Z^{\overline{\rm MS}}_{\rm DV}/Z^{\rm RI'}_{\rm DV2}$.
This requirement for different conversion factors results from the
fact that the Z-factors in the continuum ${\overline{\rm MS}}$-scheme
do not depend on the external indices, $\mu,\,\nu$ (see Eq. (2.5) of
Ref.~\cite{Gracey:2003mr}), while the results in the RI$'$-MOM scheme
do depend on  $\mu$ and $\nu$. We also need another factor $R(2
GeV,\mu)$ that will bring all Z-factors down to $\mu=2$ GeV, for example
\be
Z^{\overline{\rm MS}}_{\rm DV1}(2 GeV)=R_{\rm DV}(2 GeV,\mu)\cdot C_{\rm DV1}(\mu)\cdot Z^{\rm RI'}_{\rm DV1}(\mu)
\ee

\begin{figure}[h]
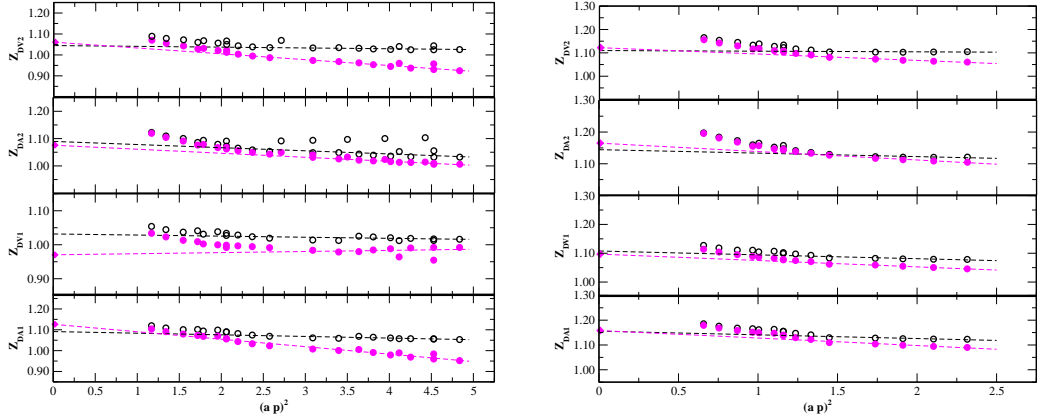

\centerline{\psfig{figure=Z_oneD_MSbar_2GeV_b3.9_0.0085_PoS.eps,height=5.5truecm,clip=}$\qquad$\psfig{figure=Z_oneD_MSbar_2GeV_b4.20_0.0065_PoS.eps,height=5.5truecm,clip=}}
\vspace{0.1cm}
   \caption{ Renormalization factors at $\beta=3.9,\,m_\pi=0.430$~GeV
     (left panel) and $\beta=4.20,\,m_\pi=0.476$~GeV (right panel) in the
     ${\overline{\rm MS}}$-scheme at 2~GeV. The lines show
     extrapolations to $a^2p^2=0$ within the range  $p^2\sim 15-32$
     (GeV)$^2$. (Open points:
  unsubtracted, filled points: subtracted)} 
\label{fig4}
\end{figure}

A ``renormalization window'' should exist for $\Lambda_{QCD}^2 <<
\mu^2 << 1/a^2$ where perturbation theory holds and finite-$a$
artifacts are small, leading to scale-independent results
(plateau). In practice such a condition is hard to satisfy: The upper
range of the inequality is extended to $(2-5)/a^2$ leading to lattice artifacts in
our results that are of ${\cal O}(a^2p^2)$. Fortunately our
perturbative calculations allow us to subtract the leading
perturbative $O(a^2)$ lattice artifacts which alleviates the
problem. To remove the remaining $O(a^2p^2)$ artifacts we extrapolate
linearly to $a^2p^2=0$ as demonstrated in Fig. \ref{fig4}.
The statistical errors are negligible and therefore an estimate of the
systematic errors is important. We note that, in general, the evaluation
of systematic errors is difficult. The largest systematic error comes
from the choice of the momentum range to
use for the extrapolation to $a^2p^2=0$. One way to estimate
this systematic error is to vary the momentum range where we perform
the fit. Another approach is to fix a range and then eliminate a given
momentum in the fit range and refit. The spread of the results about
the mean gives an estimate of the systematic error. In the final
results we give as systematic error the largest one from using these
two procedures which is the one obtained by modifying the fit
range. In order to treat all beta values equally, we fix the momentum
range in physical units and we thus fit all renormalization constants
in the same physical momentum range, $p^2\sim 15-32$ (GeV)$^2$. The
momentum interval in physical units has bean chosen such as a good
plateau exists at each $\beta$, as can be seen in Fig.~\ref{fig4}. 
The ${\cal O}(a^2)$ perturbative terms
which we subtract, decrease as $\beta$ increases, as expected.
The momentum range in lattice units at each $\beta$ is
rescaled as follows: $\beta=3.9: a^2p^2\sim 3-5$, $\beta=4.05:
a^2p^2\sim 1.9-3$, $\beta=4.20: a^2p^2\sim 1.2-2.5$. 
Our results for the ${\cal O}(a^2)$ corrected $Z$-factors in
the $\overline{\rm MS}$-scheme at 2~GeV are given in Table~\ref{tab7},
which have been obtained by extrapolating linearly in $a^2p^2$. For
$Z_{\rm DV}$ and $Z_{\rm DA}$ we used the fixed momentum range
$p^2\sim 15-32$ (GeV)$^2$~\cite{ACKPF1}, while for $Z_{\rm V}$ and $Z_{\rm A}$ we
used all the data points available, since the plateau is good for
all momenta. The final results for $Z_{\rm V}$ and $Z_{\rm A}$ for a
more extended momentum range will appear in~\cite{ACKPF2}.
{\small{
\begin{table}[h]
\begin{center}
\begin{minipage}{15cm}
{\small{
\begin{tabular}{lr@{}lr@{}lr@{}lr@{}lr@{}lr@{}l}
\hline
\hline\\[-2.25ex]
\multicolumn{1}{c}{$\beta$}&
\multicolumn{2}{c}{$\,\,Z_{\rm V}\,\,$} &
\multicolumn{2}{c}{$\,\,Z_{\rm A}\,\,$} &
\multicolumn{2}{c}{$\,\,Z_{\rm DV1}\,\,$} &
\multicolumn{2}{c}{$\,\,Z_{\rm DV2}\,\,$} &
\multicolumn{2}{c}{$\,\,Z_{\rm DA1}\,\,$} &
\multicolumn{2}{c}{$\,\,Z_{\rm DA2}\,\,$} \\
\hline
\hline\\[-2.25ex]
3.90  &$\,\,0$.&6343(6)(3)   &$\,\,0$.&7561(6)(5)  &$\,\,$&0.970(34)(26) &$\,\,$&1.061(23)(29)   &$\,\,$&1.126(22)(78)    &$\,\,$&1.076(5)(1) \\ 	 
4.05  &$\,\,0$.&6628(7)(14)  &$\,\,0$.&7722(6)(3)  &$\,\,$&1.033(11)(14) &$\,\,$&1.131(23)(18)   &$\,\,$&1.157(9)(7)      &$\,\,$&1.136(5)     \\   
4.20  &$\,\,0$.&6854(5)(13)  &$\,\,0$.&7870(5)(9)  &$\,\,$&1.097(4)(6)   &$\,\,$&1.122(7)(10)    &$\,\,$&1.158(7)(7)      &$\,\,$&1.165(5)(10) \\
\hline
\hline
\end{tabular}}}
\end{minipage}
\end{center}
\vspace{-0.35cm}
\caption{Renormalization constants in
  the ${\overline{\rm MS}}$ scheme, after extrapolating linearly in
  $a^2p^2$. The error in the first parenthesis is statistical and the
  one in the second parenthesis is systematic.} 
\label{tab7}
\end{table}
}}
\section{Conclusions}
\label{sec6}
The values of the renormalization factors for the one-derivative
twist-2 operators are calculated non-perturbatively. The method of
choice is to use a momentum dependent source and extract the
renormalization constants for all the relevant operators, which leads
to a very accurate evaluation of these renormalization  factors using
a small ensemble of gauge configurations. 
We studied the quark mass dependence and found that an extrapolation to
zero quark mass changes the result by about 1 per mille for all the
operators we presented here. This is in most cases by an order of magnitude smaller
than the systematical errors due to lattice artifacts, therefore a
calculation at a single quark mass suffices.
For all the
renormalization constants shown here we do not find any
light quark mass dependence within our small statistical
errors. Therefore it suffices to calculate renormalization constants at a given quark
mass. Despite using lattice spacing smaller than
1~fm, ${\cal O}(a^2)$ effects are sizable, thus, we perform a perturbative
subtraction of ${\cal O}(a^2)$  terms. This leads to a smoother
dependence of the renormalization constants on the momentum values at
which they are  extracted. Residual ${\cal O}(a^2p^2)$ effects are
removed by extrapolating to zero. In this way we can accurately
determine the renormalization constants in the RI$'$-MOM scheme. In
order to compare with experiment we convert our values to the $\rm
\overline{MS}$ scheme at a scale of 2~GeV. The systematic errors are estimated by
ochanging the window of values of the momentum used to extrapolate to
$a^2p^2=0$.

\end{document}